\documentclass[letterpaper]{article} 
\usepackage{aaai2026}  
\usepackage{times}  
\usepackage{helvet}  
\usepackage{courier}  
\usepackage[hyphens]{url}  
\usepackage{graphicx} 
\usepackage{natbib}  
\usepackage{caption} 
\usepackage{algorithm}
\usepackage{algorithmic}
\usepackage{latexsym}
\usepackage{multirow}
\usepackage{comment}
\usepackage{inconsolata}
\usepackage{booktabs}
\usepackage{graphicx}
\usepackage{amsfonts}
\usepackage{amsmath}
\usepackage{minted}
\usepackage{listings}
\usepackage{xcolor}
\usepackage{newfloat}
\usepackage{listings}
\DeclareCaptionStyle{ruled}{labelfont=normalfont,labelsep=colon,strut=off} 
\floatstyle{ruled}
\newfloat{listing}{tb}{lst}{}
\floatname{listing}{Listing}
\title{Automated Creation and Enrichment Framework for Improved Invocation of Enterprise APIs as Tools}

\author {
    Prerna Agarwal,
    Himanshu Gupta,
  Soujanya Soni,
  Rohith Vallam,\\
  Renuka Sindhgatta,
  Sameep Mehta
}
\affiliations {
    IBM Research, India
}
\usepackage{bibentry}

\begin{document}

\maketitle

\begin{abstract}
Recent advancements in Large Language Models (LLMs) has lead to the development of agents capable of complex reasoning and interaction with external tools.
In enterprise contexts, the effective use of such tools that are often enabled by application programming interfaces (APIs), is hindered by poor documentation, complex input or output schema, and large number of operations. These challenges make tool selection difficult and reduce the accuracy of payload formation by up to 25\%.
We propose ACE, an automated tool creation and enrichment framework that transforms enterprise APIs into LLM-compatible tools. ACE, (i) generates enriched tool specifications with parameter descriptions and examples to improve selection and invocation accuracy, and (ii) incorporates a dynamic shortlisting mechanism that filters relevant tools at runtime, reducing prompt complexity while maintaining scalability. We validate our framework on both proprietary and open-source APIs and demonstrate its integration with agentic frameworks. To the best of our knowledge, ACE is the first end-to-end framework that automates the creation, enrichment, and dynamic selection of enterprise API tools for LLM agents.
\end{abstract}


\section{Introduction}
Recent advancements in large language models (LLMs) has significantly improved their capabilities for reasoning over complex tasks and interacting with external tools, permitting the development of agents~\cite{aisurvey2025}. A central paradigm in enabling actions is \textit{tool learning}, that aims to combine the strengths of specialized tools and LLMs, thus extending their capabilities beyond text generation~\cite{toollearning2024}. Existing research on tool learning focuses on improving the agent's ability to decompose tasks and select tools, typically through supervised fine-tuning on tool-use datasets~\cite{toolformer,apibank}. However, an equally critical aspect is the tool interface as an  agent's performance is constrained not only by its reasoning ability but also by the semantic information presented by the tools. 

In many enterprise settings, tools take the form of application programming interfaces (APIs) to perform tasks. These APIs are typically designed for use by developers, with technical specifications and parameters that are not immediately suitable for invocation by LLM-based agents. Integrating enterprise APIs as tools for LLM-based agents presents several challenges. First, converting API specifications into agent-compatible tools is often a manual, time-consuming, and error-prone process, further complicated by the need to support diverse agentic frameworks. Second, enterprise APIs often lack structured, detailed, and usable documentation. This includes tool-level descriptions and parameter semantics. Lack of details limits the language agent's ability to (i) select the appropriate tool for a given natural language query and (ii) construct valid, schema-compliant input payloads. The problem is compounded by complex, nested input/output schema, which are difficult for agent's to interpret without explicit guidance or examples. Third, APIs with large operation sets; (for example; Jira, a service management platform, has over 900 API operations\footnote{\url{https://developer.atlassian.com/cloud/jira/service-desk-ops/rest/v2/intro/}}) result in scalability issues,  as passing all tool details may exceed the LLM’s context length during tool invocation. 

To address these challenges, we propose an automated enrichment framework that enriches API specifications, and a tool creation framework that transforms APIs as Python tools and augments tool definitions with structured, model-friendly docstrings. The framework parses existing schema definitions to generate clear tool-level descriptions, parameter-level documentation with type information, and illustrative example values, including concise representations of complex structures. These example values act as few-shot demonstrations, enabling the LLM to form valid input payloads. By providing enriched docstrings with sufficient context and examples, the framework supports more reliable tool selection, accurate input construction, and scalable API integration for LLM-based agents. Additionally, a shortlisting component dynamically selects the top-$k$ relevant tools for a given user query, reducing the candidate set and improving selection accuracy. Together, these capabilities improve tool selection, input construction, and scalability when integrating large and complex APIs into LLM-based agents.

\noindent\textbf{Contributions:} To summarize, we address the above-discussed challenges and make the following contributions in our work:

(1) We propose \textbf{Automated Creation and Enrichment (ACE) framework}, end-to-end system for automated creation, enrichment, and shortlisting of enterprise API tools for LLM-based agents.

(2) We introduce an automated creation process that is framework-agnostic and an enrichment process that augments tool metadata with detailed descriptions, parameter information, and illustrative examples to support accurate tool selection and calling.

(3) We present a shortlisting mechanism that identifies the top-$k$ most relevant tools for a given user query, enabling efficient and accurate tool usage in large tool repositories.

(4) We evaluate ACE framework on both proprietary and open-source enterprise APIs, demonstrating its effectiveness and seamless integration with existing agentic frameworks.

To the best of our knowledge, ACE framework is the first one to automate the API tool creation and enrichment in an end-to-end manner with tool shortlisting capability. 

\section{Related Work}
We present existing work in the context of Tool learning: Tool Creation, Usage, and Shortlisting~\cite{toollearning2024}.

\paragraph{API Tool Creation}
Recent work such as LangChain's \texttt{OpenAPI Toolkit}\footnote{\url{https://python.langchain.com/docs/integrations/tools/openapi/}} enables LLMs to interact with APIs by parsing the OpenAPI schema and allowing the agent to explore the API documentation at runtime. Rather than exposing individual endpoints as structured tools, this approach relies on intermediate reasoning over the Open API Specification (OAS). While effective for smaller APIs, enterprise APIs often include hundreds of operations and large, complex schemas, which can exceed the context capacity of smaller models and reduce scalability. 

Hence, transforming an OAS into framework-specific Python tools, is essential for enabling seamless integration with LLM-based agents. The proposed framework implements a transformation pipeline that parses the OAS, infers operations, input and output schemas, and generates framework-specific Python tool definitions, allowing a subset of tools to be imported into an agentic framework for real-time interaction.

\paragraph{Tool Usage}
Recent work has focused on improving Large Language Models’ (LLMs) ability to interact with external APIs and tools. Tool usage research has progressed through model-side improvements: (i) generating tool calling data and fine-tuning LLMs~\cite{gorilla, toolllm}, and (ii) benchmarking design to evaluate tool selection, sequencing, and invocation under realistic constraints~\cite{taubench}. However, enhancing the effectiveness of LLMs in using tools also requires improving how tools are constructed and presented to the model.

To address limitations in tool specifications, OASBuilder \cite{OASBuilder} generates OpenAPI schemas from webpages and adds missing endpoint descriptions, parameter documentation, and constraints (e.g., formats, enum values). This improves the completeness of the OAS, which is critical for generating effective tool representations. EasyTool \cite{EasyTool} extracts tool descriptions and guidelines from diverse and often inconsistent documentation sources, helping LLMs understand tool purpose and usage more reliably. ToolCoder~\cite{ToolCoder} takes a generative approach, producing both Python tool code and natural language descriptions from a user query.

While these methods either generate an OAS or a Python tool, they do not specifically address the challenges of  transforming large, complex OAS into agent-usable tools at enterprise scale. Our work complements these efforts by proposing an automated OAS-to-tool framework that integrates creation, enrichment, and shortlisting to support accurate and scalable tool use in LLM-agent systems.
\paragraph{Tool Shortlisting}
Early approaches rely on the LLM agent choose from static tool lists or on rule-based mappings, which are difficult to scale and prone to failure when handling ambiguous, under-specified, or multi-intent queries. To address these limitations, recent work has explored \textit{retrieval-augmented strategies} that leverage semantic search to dynamically shortlist tools based on natural language task descriptions. For example, RAG Agents combine dense vector retrieval with LLM-based reasoning to improve tool selection precision~\cite{karia2024ragagents}. More recent methods such as MCP-Zero and ScaleMCP demonstrate that semantic retrieval pipelines can scale to thousands of tools while reducing prompt size and improving retrieval accuracy~\cite{jiang2025mcpzero,zhang2025scalemcp}. These studies collectively demonstrate effectiveness of embedding-based retrieval for dynamic tool shortlisting. Our framework explicitly utilizes the enriched tool metadata to shortlist tools.
\begin{figure*}
    \centering
    \includegraphics[width=0.8\linewidth]{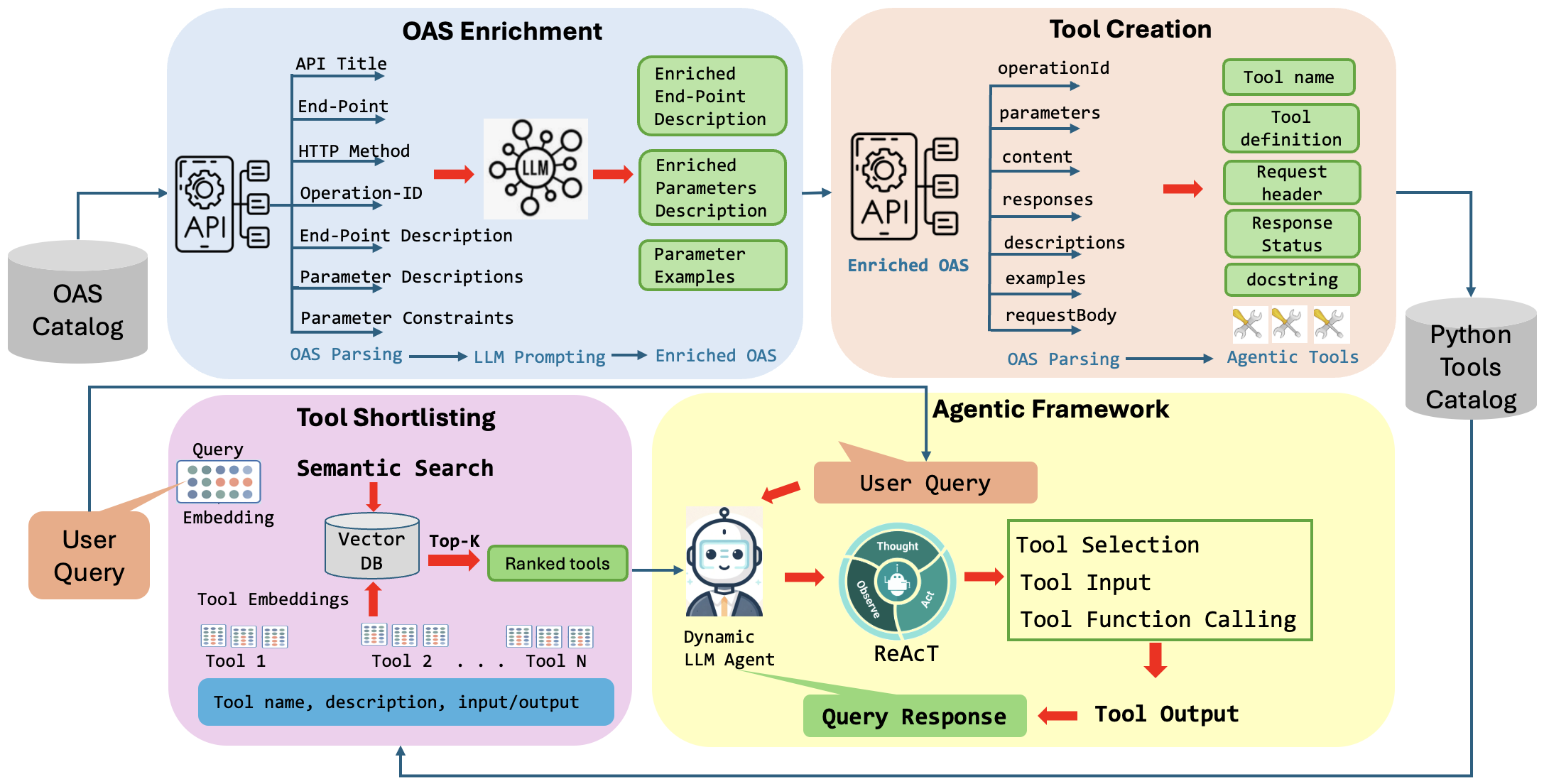}
    \caption{Automated Creation and Enrichment (ACE) Framework}
    \label{fig:framework}
    \vspace{-0.3cm}
\end{figure*}

\section{Automated Creation and Enrichment (ACE) Framework}
The Automatic Creation and Enrichment (ACE) Framework for Enterprise APIs as Tools is shown in Figure \ref{fig:framework}. All the APIs present in the catalog first undergo `Enrichment', and then are transformed into the Agentic Framework specific Python tools. The final enriched tools gets stored in the tool catalog. 
Given a user query, the relevant top-$k$ tools are shortlisted and loaded dynamically into the Agentic Framework, if shortlisting is enabled. If not, the entire tool catalog is loaded. The agent serves the user query by selecting and invoking a tool thereby forming appropriate tool input and obtaining the tool output. 
\subsection{OAS Metadata Enrichment}
This section describes a method for enhancing OAS metadata by exploiting the internal structure of OAS documents. Although OAS is primarily intended to provide a machine-readable description of RESTful interfaces, its components often encode implicit semantic information. Elements such as endpoint paths, HTTP methods, operation IDs, API titles, parameter definitions, and schema constraints can be used to infer an endpoint’s function, the nature of its inputs and outputs, and typical usage scenarios. By making these inferences explicit, we can produce richer metadata that supports more effective use of APIs by automated agents.

\begin{itemize}
\item{\textbf{API Method Description:}} If an API endpoint's description is missing, inadequate, or unclear, we generate a new one based on contextual information surrounding the endpoint. Elements such as the HTTP method (GET, POST, etc.) and operation ID often reveal the method's purpose and functionality, while the parameter section outlines the expected inputs. By synthesizing these components, we can produce richer and more accurate method descriptions that clearly convey both the intent of the endpoint and the required input data.

\item{\textbf{Parameter Descriptions:}} Similarly, when parameter descriptions are missing or vague, we analyze contextual cues such as the API method name, parameter name, and operation ID to infer the parameter’s meaning, intent, and purpose. Additionally, the OAS may explicitly define constraints—for example, allowed enum values or required input formats. By synthesizing this information, we generate concise and informative parameter descriptions that clearly convey the parameter’s role, any constraints, and how it should be used. 

\item{\textbf{Parameter Examples:}} In addition to generating parameter descriptions, we also use the OAS structure to derive meaningful parameter examples. We use various details - parameter name, type and description, operation ID etc to generate parameter examples. The updated parameter description is used for this purpose. If the parameter description mentions any parameter constraints, the generated examples will be consistent with these constraints. 
\end{itemize}

To support OAS enrichment, we design dedicated LLM prompts for each of the three metadata generation tasks described above. The prompts incorporate the relevant instructions and contextual signals outlined earlier, guiding the LLMs to produce accurate and informative metadata. Given an OAS, we first parse it to extract individual endpoints and their surrounding context. We then identify the key constructs like API title, operation ID, end-point path etc and invoke the appropriate LLM prompt for each task with the relevant inputs.

Figure~\ref{fig:enrich3-oas} (Appendix) shows a representative OAS. As shown at the left side, the end-point contains a very basic end-point description - ``delete a LimitRange''. The right side has the enriched OAS, where the tool description, parameter descriptions are generated along-with examples. 
\subsection{API Tool Creation}
We assume the OAS as input for generating Python tools that can be used in LLM agent frameworks. OAS \cite{oapi2023} is the industry standard for defining RESTful API interfaces. 
A tool is created by parsing different components of each path operation or endpoint in an OAS as follows (see Figure~\ref{fig:enrich1-tool} in Appendix):
\begin{enumerate}
\item \textit{Imports and Tool Decorator:} Framework-specific libraries are imported to enable tool creation. The tool decorator is then applied to the function, marking it as callable by the LLM.
\item \textit{Function Definition:} The \texttt{operationId} specified in the OAS is used as the function name. Path parameters, query parameters, and request body fields are extracted along with their metadata (e.g., required/optional status, data types) to define function arguments.
\item \textit{Headers:} Request headers are derived from the OAS \texttt{content} field. When authentication is required, credentials are retrieved from the configuration file and included in the request.
\item \textit{Docstring:} To provide the LLM with complete tool documentation, the docstring is generated from: (1) endpoint descriptions, (2) parameter descriptions, and (3) an illustrative input example with sample parameter values. The example serves as a one-shot prompt, helping the LLM generate correct payload while reducing hallucinations.
\item \textit{API Request Construction:} Input arguments defined in the function are mapped to their corresponding positions in the request (path, query string, or body). This mapping ensures that the final API call conforms to the endpoint specification.
\item \textit{Response Handling:} The \texttt{responses} field in the OAS is parsed to enumerate possible status codes. At runtime, the API returns responses according to execution status, which are surfaced as tool outputs.
\end{enumerate}

\subsection{Tool Shortlisting}
As agents interact with large, diverse toolsets, tool shortlisting becomes essential. Without it, agents needs to consider the entire toolset $T$, comprising hundreds of functions, making it impractical to fit into a prompt or perform real-time reasoning and execution. Shortlisting narrows this space semantically, preserving relevance while enabling scalable decisions and more accurate tool use in practice.

Given the user query $q$, the agent retrieves a ranked list of $k$ tool candidates $\{T_1, T_2, \ldots, T_k\}$ from a tool catalog $\mathcal{T}$, where each tool $T_i \in \mathcal{T}$ is represented by metadata such as name, description and input/output schemas. The challenge lies in mapping the latent semantic meaning of $q$ to appropriate tool affordances under varying degrees of ambiguity, underspecification, or multi-intent composition.

We employ a RAG-based shortlisting approach, that leverage semantic vector representations to address the mapping problem. Both the user intent $q$ and tool descriptions $d_i \in \mathcal{D}$ are embedded into a shared vector space $\mathbb{R}^n$ using Sentence Transformer\footnote{\url{https://huggingface.co/sentence-transformers/paraphrase-MiniLM-L3-v2}}. However, other sentence-level encoders such as BGE-M3, OpenAI Ada v2 can also be used here. The shortlisting then proceeds via approximate nearest neighbor (ANN) search to identify high-similarity matches under cosine or Euclidean distance. Formally,
\[
T_{\text{shortlisted}} = \text{Top}_k\left(\text{sim}(E(q), E(d_i)) \ \forall \ d_i \in \mathcal{D}\right)
\]
where $E(\cdot)$ is the embedding function, and $\text{sim}$ is the similarity metric.
We expect the enriched metadata to provide similar and relevant tools while shortlisting.

\subsection{Tool Execution}
We use LangChain as the base agentic framework and the standard ReAct agent for experimentation~\cite{react2023}. The ReAct pattern combines reasoning and action in an iterative loop: the agent reasons about the task, invokes a tool, observes the outcome, and repeats as needed. Our framework remains pluggable with other agentic frameworks and agent patterns.

Each interaction begins with a natural language user query. The agentic environment is then initialized with the top-$k$ tools selected by the shortlister, or with all tools in the catalog if shortlisting is disabled. The agent chooses the appropriate tool from the available set, executes it with the required input, and retrieves the tool output.

\section{Experiments}
We enrich the API tools in three different variants and benchmark the below against the defined metrics.
\begin{itemize}
\item {\textbf{No Enrich:}} Here the original tool description  is used added in the tool docstring.
\item {\textbf{Enrich-1:}} Here only tool description is enriched and added in the tool docstring.
\item {\textbf{Enrich-2:}} Here the tool description and parameter descriptions are enriched and added in the tool docstring.
\item {\textbf{Enrich-3:}} Here the generated parameter examples are also added along with enriched tool description and parameter descriptions in the tool docstring. 
\end{itemize}


We evaluate the proposed ACE framework by answering the following Research Questions (RQs):
\begin{itemize}
    \item \textbf{RQ1:} How does the metadata created from enriched OAS affect the tool calling performance?
    \item \textbf{RQ2:} How are different LLMs in the agentic framework impacted by the metadata for tool selection and calling?
    \item \textbf{RQ3:} How does the metadata created from enriched OAS affect the tool shortlisting?
\end{itemize}
\subsection{Datasets and Models}
\noindent\textbf{APIs}: We evaluate ACE framework on the following proprietary (Salesloft) and open-sourced (Kubernetes) enterprise APIs each with one operation and endpoint:
\begin{itemize}
    \item \textbf{Salesloft\footnote{\url{https://developers.salesloft.com/docs/api/}}:} These are enterprise sales engagement APIs with different levels of complexity in terms of input parameters, response output parameters.
    \item \textbf{Kubernetes\footnote{\url{https://kubernetes.io/docs/concepts/overview/kubernetes-api}}:} APIs enables query and manipulation of the state of objects in Kubernetes such as Pods, Namespaces, ConfigMaps, and Events. Comparatively, this is a more complex dataset with various tools being similar in names and having related functionalities. While number of parameters are less, the request body is complex with nested schema.
\end{itemize} 
Dataset statistics are shown in Table \ref{tab:dataset}. 
\begin{table}[]\small
    \centering
    \begin{tabular}{p{1.2cm}p{0.7cm}cc}
    \toprule
      \textbf{Dataset}  & \textbf{\#APIs} & \textbf{\#Input Params}& \textbf{\#NL Utterances}\\
      \midrule
        Salesloft & 42&0-16&130\\
        Kubernetes & 86 &0-7&164\\
    \toprule
    \end{tabular}
    \caption{Dataset Statistics}
    \label{tab:dataset}
    \vspace{-0.4cm}
\end{table}

\noindent\textbf{User Query (NL Utterance):} We simulate complex user queries by generating NL utterances for each dataset using an LLM. Each NL utterance is then manually corrected to resemble real-world enterprise user query.

\noindent\textbf{Models:} We use different open-source LLMs of varying size to evaluate ACE framework i.e., \texttt{granite3.3-8b-instruct} (Granite-8B)\footnote{\url{https://huggingface.co/ibm-granite/granite-3.3-8b-instruct}}, \texttt{llama3.3-70b-instruct} (Llama-70B)\footnote{\url{https://huggingface.co/meta-llama/Llama-3.3-70B-Instruct}}, \texttt{llama3.1-405b-instruct} (Llama-405B)\footnote{\url{https://huggingface.co/meta-llama/Llama-3.1-405B-Instruct}}. We use \texttt{mistral-large}\footnote{\url{https://huggingface.co/mistralai/Mistral-Large-Instruct-2407}} to generate NL utterances. We do not evaluate using close-sourced models such as GPT-4 due to the costs and deployment aspects associated with them.
\subsection{Metrics} 
Following metrics are used to evaluate ACE performance:
\begin{itemize}
    \item \textbf{Tool Selection Accuracy (S):} It is computed as the fraction of NL utterances for which the correct tool was invoked by the agent.
    \item \textbf{Tool Calling Input Errors: } It is computed as the fraction of tool input parameters that were not formed correctly by the LLM. We report this metric only for the NL utterances for which the right tool was selected by the agent. This metric is computed in 3 dimensions: (a) \textbf{type mismatch (T)}: fraction of tool input parameters where the type of input parameter is incorrect - the tool definition expects a type but the LLM passes a different type; (b) \textbf{missing parameters (M)}: fraction of tool input parameters where the input parameter is present in the NL utterance but missing from the tool input; (c) \textbf{incorrect parameters (I)}: fraction of tool input parameters that are incorrect/hallucinated by the LLM, not appearing in the NL utterance.
\end{itemize}

\begin{table*}[t]\scriptsize
    \centering
    \begin{tabular}{ccp{0.3cm}p{0.5cm}p{0.5cm}p{0.55cm}|p{0.35cm}p{0.45cm}p{0.5cm}p{0.5cm}|p{0.35cm}p{0.45cm}p{0.5cm}p{0.5cm}|p{0.35cm}p{0.45cm}p{0.5cm}p{0.5cm}}
    \toprule
        \multirow{2}{*}{\textbf{Dataset}} & \multirow{2}{*}{\textbf{Models}} &\multicolumn{4}{c}{\textbf{No Enrich}} & \multicolumn{4}{c}{\textbf{Enrich-1}}& \multicolumn{4}{c}{\textbf{Enrich-2}}& \multicolumn{4}{c}{\textbf{Enrich-3}}\\
        \cline{3-18}
        & & S\% & T\% & M\% & I\% & S\% & T\% & M\% & I\% & S\% & T\% & M\% & I\% & S\% & T\% & M\% & I\%\\
        \midrule
        \multirow{3}{*}{Salesloft} & Granite-8B & 96.1 & 0.15 & 6.1 & 3.4 & 97.0 & 0.0 & 15.0 & 8.7 & 96.1 &0.13 & 4.8& 3.0 & 96.1& 0.0 & 0.8 & 1.5\\
        & \multirow{1}{*}{Llama-70B} & 100 & 23.4 & 6.6 & 4.0 &   100 & 23.5 & 9.1 & 5.6 &97.7 &15.9 & 5.9 & 3.3 &99.2  &  16.9& 0.1 &0.55\\
         & \multirow{1}{*}{Llama-405B} & 100 & 21.7 & 43.3 & 0.4 & 100 & 24.2 & 36.3 & 0.13 & 100 &20.72 & 44.0& 0.14 & 100 & 25.2 & 41.3 & 0.26\\
         \midrule
         \multirow{3}{*}{Kubernetes} & Granite-8B & 71.8  & 0.3 & 9.4   & 5.3 &  59.2& 0.0 & 12.3 & 5.3  & 65.3 & 0.0 & 9.2 & 1.7  & 65.3 & 0.0 & 12.2  & 8.0 \\
         & \multirow{1}{*}{Llama-70B}& 90.3 & 11.4  & 0.95  & 19.1  & 93.2 & 13.9  & 0.9  & 18.9  & 92.7  & 7.3 & 0.7 & 13.4  & 91.2 & 3.5 & 0.7  & 12.0 \\
         & \multirow{1}{*}{Llama-405B}& 91.3 & 12.3 & 2.5 & 11.3 &89.5  & 15.3 & 3.0 & 15.3  & 90.8  & 14.9 & 2.2& 12.2 & 92.0 & 15.1 & 1.3 & 13.7 \\
    \toprule
    \end{tabular}
    \caption{Results on Kubernetes and Salesloft Datasets}
    \label{tab:enrichment}
\end{table*}


\begin{table}[t]\scriptsize
\centering
\begin{tabular}{llccccc}
\hline
\textbf{Dataset} & \textbf{Method} & \textbf{Top 3} & \textbf{Top 5} & \textbf{Top 10} & \textbf{Top 15} & \textbf{Top 20} \\
\hline
\multirow{4}{*}{Salesloft} 
& No Enrich & 89.23 & 91.54 & 96.15 & 96.15 & 96.15\\
& Enrich-1  & 90.00 & 94.62 & 96.15 & 96.15 & 96.15 \\
& Enrich-2  & 90.00 & 93.85 & 95.38 & 95.38 & 96.92 \\
& Enrich-3  & 88.46 & 93.85 & 93.85 & 93.85 & 93.85 \\
\hline
\multirow{4}{*}{Kubernetes} 
& No Enrich & 71.34 & 82.32 & 90.85 & 94.51 & 96.95 \\
& Enrich-1  & 77.44 & 87.20 & 92.68 & 96.34 & 98.17 \\
& Enrich-2  & 81.71 & 86.59 & 92.68 & 95.73 & 98.17 \\
& Enrich-3  & 80.49 & 86.59 & 91.46 & 95.73 & 98.17 \\
\hline
\end{tabular}
\caption{Accuracy (in \%) of Tool Shortlisting}
\label{tab:shortlist}
\vspace{-0.4cm}
\end{table}

\section{Results and Discussion}
We discuss the results  on both the datasets with ablations on enrichment and shortlisting.
\subsection{RQ1: Impact of Enrichment on Tool Calling}

Table~\ref{tab:enrichment} present the results of the enrichment experiments on the two datasets. 

\paragraph{Tool Selection Accuracy (S):}
For the Salesloft dataset, tool selection accuracy is already saturated (close to 100\%) across all enrichment variants, leaving little room for improvement. For Kubernetes, however, performance varies significantly across models and enrichments. With Enrich-1, Llama-70B improves by +2.9 points, while Llama-405B shows only marginal gains. For the Granite model, the no-enrichment variant performs best, outperforming enriched variants by about 5\%, suggesting that smaller models may be confused by additional verbosity. There is an approximate 10\% difference in average accuracy between Salesloft and Kubernetes. This gap is primarily due to the fact that, as the number of tools in the catalog increases, the LLMs' ability to disambiguate between them tends to decrease.

\paragraph{Tool Input Calling Errors:}
Enrichment generally improves parameter correctness, especially with Enrich-3.

\textbf{Type mismatch errors (T)} for both Salesloft and Kubernetes datasets, are almost zero for Granite-8B across all enrichments. For Llama-70B, significant reductions occur with Enrich-2 and Enrich-3, with Enrich-3 outperforming Enrich-2. In contrast, Llama-405B does not leverage extra metadata effectively, as it often considers all parameters of string type leading to persisting type mismatches.

\textbf{Missing parameter errors (M)} are reduced with Enrich-2 and Enrich-3, as parameter descriptions help, and Enrich-3 further lowers errors by including request body examples. For Llama-405B, type errors often propagate into missing parameter errors. Smaller models such as Granite-8B struggle on Kubernetes, where verbosity and similar parameter names create confusion.

\textbf{Incorrect parameter errors (I)} follow a similar trend. Enrich-3 yields significant improvements for Granite-8B and Llama-70B on Salesloft, and also benefits Llama-70B significantly on Kubernetes.

An example of how enrichment helps in correct tool input formation is shown in Figure \ref{fig:example-enrichment} (Appendix).

\vspace{10pt}
\noindent\fbox{%
    \parbox{0.45\textwidth}{%
    \textbf{Summary (RQ1):} Effective enrichment is stage-dependent — concise metadata helps agents select tools efficiently, while detailed metadata ensures reliability in forming the right input for tool calling. 
    }%
}

\subsection{RQ2: Impact of Enrichment on Agent LLM}
From our experiments on the three LLM models, we can make the following observations:

\paragraph{Small Size Model – Granite-8B:} On Salesloft, Enrich-3 yields clear improvements in tool input calling for missing and incorrect parameters. However, in Kubernetes, tool selection degrades under Enrich-3, and input errors show no improvement, indicating the model’s sensitivity to metadata verbosity with complex schema and a large number of tools.

\paragraph{Medium Size Model – Llama-70B:} On Salesloft, Enrich-3 consistently reduces all tool input errors. In Kubernetes, missing parameters stabilize and incorrect parameters reduce (–7.1 points). We observe that the model often formats inputs incorrectly despite identifying the right parameters while considering the enrichment information. For example, instead of passing \texttt{\{'dryRun':"All", 'fieldValidation':"Ignore"\}} as strings, it produces dictionaries such as \texttt{\{'dryRun': \{'type': 'string', 'value': 'All'\}, ...\}}.

\paragraph{Large Size Model – Llama-405B:} Tool selection accuracy is near-saturated on Salesloft and Kubernetes, so enrichment provides little benefit. It worsens missing or type errors as the model overgeneralizes inputs as strings, even for numbers, booleans, or objects. This leads to unreliable tool calls, such as representing a request body as a string: \texttt{\{'requestBody': "\{'apiVersion': 'v1', 'data': \{'mute': 'True'\}"\}\}}.

\vspace{10pt}
\noindent\fbox{%
    \parbox{0.45\textwidth}{%
    \textbf{Summary (RQ2):} Results indicate that enrichment improvements are model-dependent; most beneficial for smaller models, but cannot overcome inherent model limitations such as over-generalization of types or format. 
    }%
}

\subsection{RQ3: Impact of Enrichment on Tool Shortlisting}
As shown in Table \ref{tab:shortlist}, enrichment helps to achieve better tool shortlisting performance at smaller $k$. In the Kubernetes dataset, enrichment raises Top-3 accuracy by +10\% points. Similar improvement is seen in the performance at Top-5. For larger $k$, the relevant tool is picked up irrespective of enrichment. A similar pattern holds for Salesloft, where baseline accuracy is already high, but enrichment provides modest improvements at lower $k$. Hence, enrichment is most valuable at smaller $k$, where it helps models overcome limitations of context and tool set size, enabling better tool selection. 

As smaller models are constrained by their context length, tool shortlisting helps LLM models overcome this limitation.
To investigate this, we conducted an experiment with Granite-8B (small model) where the agent dynamically obtains top-10 shortlisted tools from Enrich-3 variant. The tool selection with Enrich-3 is 66\% despite having only 10 tools provided to the agent. Note that, the performance with Enrich-3 is similar to that achieved without tool shortlisting (see Table \ref{tab:enrichment}). Hence, tool shortlisting with enrichment effectively enables LLM agents to handle large tool catalogs.

\vspace{10pt}
\noindent\fbox{%
    \parbox{0.45\textwidth}{%
    \textbf{Summary (RQ3):} Enrichment provides a boost in tool shortlisting accuracy at smaller $k$. With increase in $k$, the tool shortlisting accuracy comes closer towards 100\%.
    }%
}


\section{Deployment and Impact}

Our framework is currently deployed as part of the alpha release of IBM Watsonx Orchestrate Agent Development Kit (ADK)\footnote{\url{https://developer.watson-orchestrate.ibm.com/}}, specifically designed for the agent builder persona. The ADK provides a tooling environment where builders can configure, deploy, and manage agents and tools within Watsonx Orchestrate. The tool builders can import OAS and Python functions as tools.  The current feature would support building domain specific agents with REST APIs for a wide range of domains, including IT, human resources, finance, procurement, productivity, and sales.
Our automated enrichment feature standardizes generation of tool metadata and reduces errors caused by poor tool definitions.
\begin{figure}
    \centering
    \includegraphics[width=0.5\textwidth]{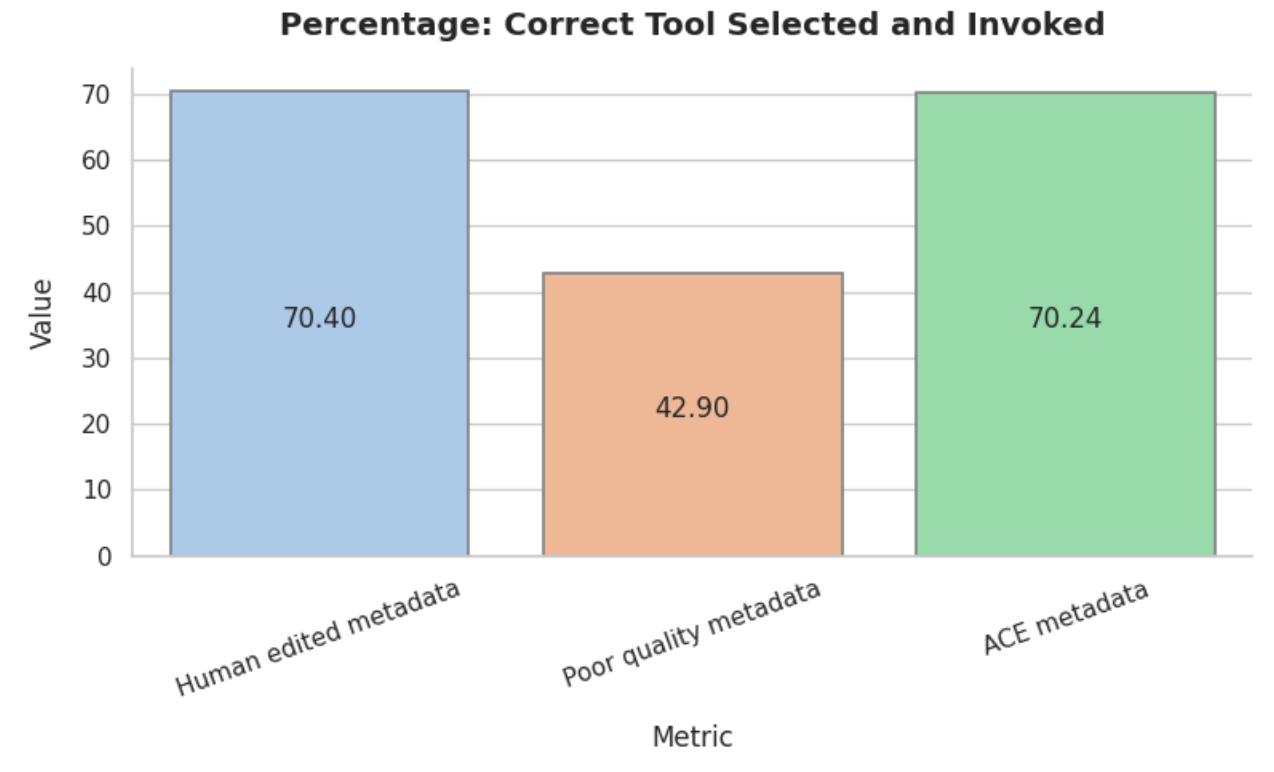}
    \caption{Evaluation on ADK Tools for IT Service Domain}
    \label{fig:wxoresult}
    \vspace{-0.4cm}
\end{figure}
We conducted internal experiments using an IT ticketing domain containing 6 agents and 46 tools developed by in-house tool builders. In this setting, tool specifications and their descriptions were carefully authored and tested manually by developers. We evaluated 3 scenarios: (i) tools with both method and parameter descriptions written by the developer, (ii) tools where the description was identical to the tool name and parameter descriptions were absent—representing a case of minimal or poor metadata, and (iii) tools where the tool name and parameter descriptions were automatically generated by the ACE framework. For each scenario, we executed 600 single-turn user utterances through the agent. The results, shown in Figure~\ref{fig:wxoresult}, report the fraction of utterances in which the correct tool was selected and successfully invoked. We observe a 27\% improvement in tool selection and invocation accuracy when metadata is enriched by ACE compared to the minimal metadata condition. Furthermore, ACE-enriched metadata achieves performance comparable to human-authored metadata.

 \section{Conclusion and Future Work}
In this work, we propose Automated Creation and Enrichment (ACE) framework that automates creation, enrichment and shortlisting of API tools. We examined tool learning from the perspective of metadata enrichment in OAS-derived enterprise tools, showing that enhanced metadata improves LLM-based agents’ ability to shortlist, select, and invoke tools effectively. Our findings highlight the importance of semantically rich tool specifications for reliable tool calling. As future work, we plan to extend this study across additional domains such as HR, finance, and procurement to assess the generality of our approach. 

\bibliography{aaai2026}

\appendix
\section{Appendix}
\subsection{Example of ACE Framework for Kubernetes}
\begin{figure*}
    \centering
    \includegraphics[width=\linewidth]{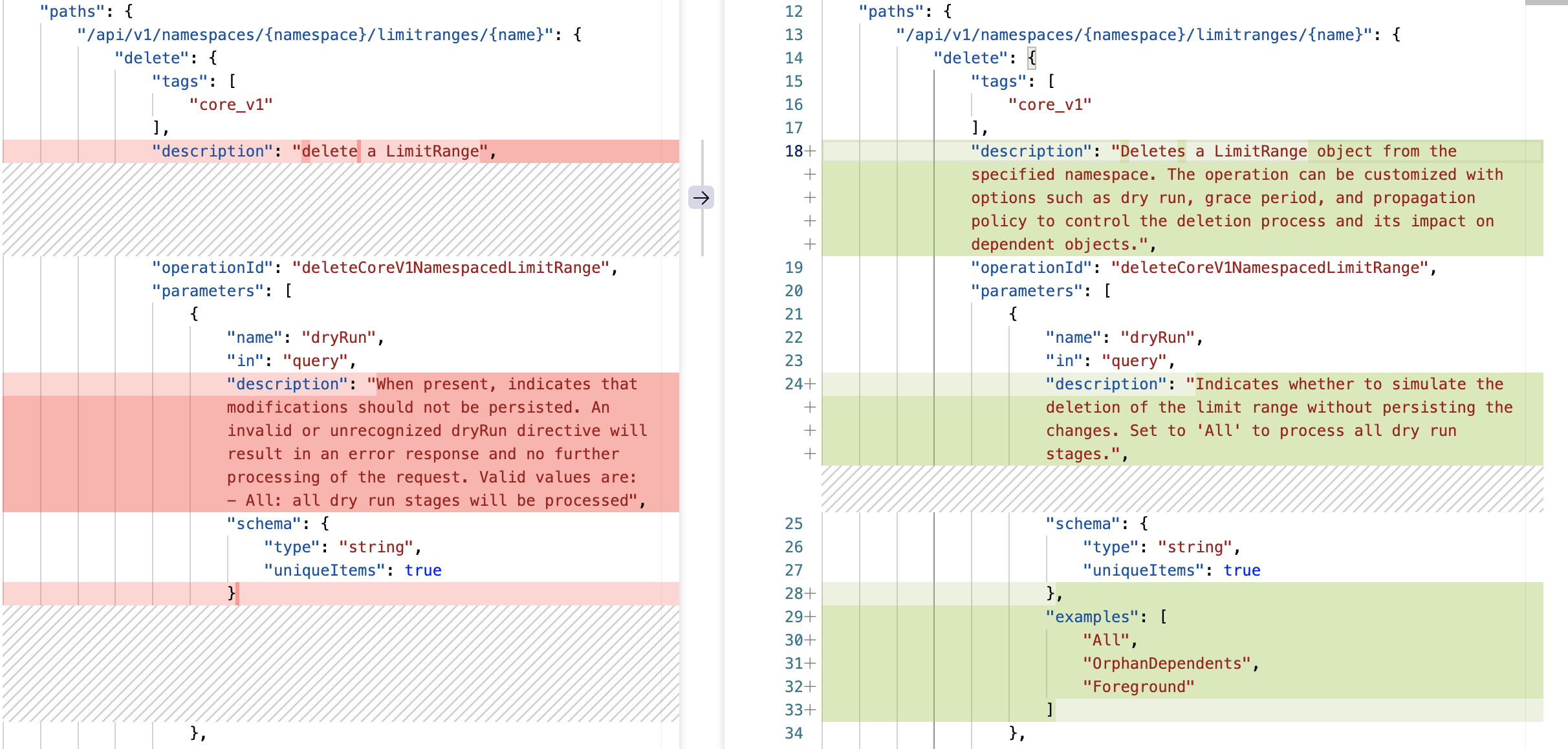}
    \caption{Enrichment of Kubernetes OAS}
    \label{fig:enrich3-oas}
\end{figure*}

Enrichment of a sample Kubernetes OAS \texttt{deleteCoreV1NamespacedLimitRange} for a parameter is shown in Figure \ref{fig:enrich3-oas}. It depicts how the tool description, parameter description text is enriched and examples are added. The tools created for different variants from the enriched OAS is shown below. As shown in Figure \ref{fig:enrich1-tool}, in Enrich-1, only the enriched tool description is presented in the docstring.  As shown in Figure \ref{fig:enrich1-tool}, in Enrich-2, the enriched tool description along with enriched parameter description is presented in the docstring.  As shown in Figure \ref{fig:enrich1-tool}, in Enrich-3, the enriched tool description, parameter description and examples are presented in the docstring.  
\begin{figure}[h]
\noindent\fbox{%
    \begin{small}
    \parbox{0.45\textwidth}{%
    \textit{NL Utterance:} ``Create a brand new endpoints resource of kind service in the namespace dev-1 by sending a POST request to the Kubernetes API. Use v1 api."\\\\
    \textit{Payload w/o enrichment}: \\ 
    \textcolor{red}{\texttt{\{`dryRun': `null', `fieldManager': `null', `fieldValidation': `null', `namespace': `dev-1', `requestBody': `\{"apiVersion": "v1", "kind": "Endpoints"\}'\}}}\\ \\
    \textit{Enriched Payload}: \\ 
    \textcolor{teal}{\texttt{\{'dryRun': None, 'fieldManager': None, 'fieldValidation': None, 'namespace': 'dev-1', 'requestBody': \{'apiVersion': 'v1', 'kind': 'Service'\}\}}}

    }%
    \end{small}
}
\caption{Kubernetes Payload Example}
\label{fig:example-enrichment}
\end{figure}

\begin{figure*}
\centering
\begin{minted}[breaklines, breakafter=d, fontsize=\small]{python}

import json
import requests
from typing import *
from langchain_core.tools import tool

@tool
def deleteCoreV1NamespacedLimitRange(namespace: str, name: str, dryRun: Optional[str] = None, gracePeriodSeconds: Optional[int] = None, orphanDependents: Optional[bool] = None, propagationPolicy: Optional[str] = None, requestBody: Optional[dict] = None):
	""" Deletes a specified LimitRange within a given namespace. This operation allows for optional parameters to control the deletion process, such as specifying a grace period before deletion, determining whether dependent objects should be orphaned, and setting a propagation policy for garbage collection. Additionally, a dry run option is available to simulate the deletion without persisting any changes.
	"""

	header = {
		'accept': 'application/json',
		'content-type': 'application/x-www-form-urlencoded'
	}
	queryParam = {'dryRun' : dryRun, 'gracePeriodSeconds' : gracePeriodSeconds, 'orphanDependents' : orphanDependents, 'propagationPolicy' : propagationPolicy}

	api_url = f"http://xxxx:8080/api/v1/namespaces/{namespace}/limitranges/{name}"
	response = requests.delete(api_url, headers=header, params=queryParam, json=requestBody)
	return {"status_code": response.status_code, "response": response.json()}

\end{minted}
\caption{Enrich-1 Python tool}\label{fig:enrich1-tool}
\end{figure*}

\begin{figure*}
\centering
\begin{minted}[breaklines, breakafter=d, fontsize=\small]{python}

import json
import requests
from typing import *
from langchain_core.tools import tool

@tool
def deleteCoreV1NamespacedLimitRange(namespace: str, name: str, dryRun: Optional[str] = None, gracePeriodSeconds: Optional[int] = None, orphanDependents: Optional[bool] = None, propagationPolicy: Optional[str] = None, requestBody: Optional[dict] = None):
	""" Deletes a specified LimitRange within a given namespace. This operation allows for optional parameters to control the deletion process, such as specifying a grace period before deletion, determining whether dependent objects should be orphaned, and setting a propagation policy for garbage collection. Additionally, a dry run option is available to simulate the deletion without persisting any changes.

	:param namespace: The namespace in which the LimitRange resource is located. This parameter is required to identify the specific namespace where the LimitRange resource to be deleted resides.
	:param name: The name of the LimitRange resource to be deleted.
	:param dryRun: Indicates whether the operation should be performed as a dry run. If set to 'All', all dry run stages will be processed. An invalid or unrecognized value will result in an error response and no further processing of the request.
	:param gracePeriodSeconds: The duration in seconds before the object is deleted. A value of zero indicates immediate deletion. If not specified, the default grace period for the object type will be used.
	:param orphanDependents: Indicates whether dependent objects should be orphaned. If set to true, the 'orphan' finalizer will be added to the object's finalizers list. If set to false, it will be removed. Note: This field is deprecated; use PropagationPolicy instead. This field will be removed in version 1.7.
	:param propagationPolicy: Specifies the garbage collection policy for the limit range. Determines whether and how dependent resources will be deleted. Acceptable values are: 'Orphan' - orphan the dependents; 'Background' - delete dependents in the background; 'Foreground' - delete all dependents in the foreground. Only one of this field or OrphanDependents can be set.
	:return: The JSON response from the API. 
	"""

	header = {
		'accept': 'application/json',
		'content-type': 'application/x-www-form-urlencoded'
	}
	queryParam = {'dryRun' : dryRun, 'gracePeriodSeconds' : gracePeriodSeconds, 'orphanDependents' : orphanDependents, 'propagationPolicy' : propagationPolicy}

	api_url = f"http://xxxx:8080/api/v1/namespaces/{namespace}/limitranges/{name}"
	response = requests.delete(api_url, headers=header, params=queryParam, json=requestBody)
	return {"status_code": response.status_code, "response": response.json()}

\end{minted}
\caption{Enrich-2 Python tool}\label{fig:enrich2-tool}
\end{figure*}

\begin{figure*}
\centering
\begin{minted}[breaklines, breakafter=d, fontsize=\small]{python}

import json
import requests
from typing import *
from langchain_core.tools import tool

@tool
def deleteCoreV1NamespacedLimitRange(namespace: str, name: str, dryRun: Optional[str] = None, gracePeriodSeconds: Optional[int] = None, orphanDependents: Optional[bool] = None, propagationPolicy: Optional[str] = None, requestBody: Optional[dict] = None):
	""" Deletes a specified LimitRange within a given namespace. This operation allows for optional parameters to control the deletion process, such as specifying a grace period before deletion, determining whether dependent objects should be orphaned, and setting a propagation policy for garbage collection. Additionally, a dry run option is available to simulate the deletion without persisting any changes.

	:param namespace: The namespace in which the LimitRange resource is located. This parameter is required to identify the specific namespace where the LimitRange resource to be deleted resides.
	:param name: The name of the LimitRange resource to be deleted.
	:param dryRun: Indicates whether the operation should be performed as a dry run. If set to 'All', all dry run stages will be processed. An invalid or unrecognized value will result in an error response and no further processing of the request.
	:param gracePeriodSeconds: The duration in seconds before the object is deleted. A value of zero indicates immediate deletion. If not specified, the default grace period for the object type will be used.
	:param orphanDependents: Indicates whether dependent objects should be orphaned. If set to true, the 'orphan' finalizer will be added to the object's finalizers list. If set to false, it will be removed. Note: This field is deprecated; use PropagationPolicy instead. This field will be removed in version 1.7.
	:param propagationPolicy: Specifies the garbage collection policy for the limit range. Determines whether and how dependent resources will be deleted. Acceptable values are: 'Orphan' - orphan the dependents; 'Background' - delete dependents in the background; 'Foreground' - delete all dependents in the foreground. Only one of this field or OrphanDependents can be set.
	:return: The JSON response from the API. 

	Input Example:
	namespace = 'default'
	name = 'default-limit-range'
	dryRun = 'All'
	orphanDependents = True
	propagationPolicy = 'Orphan'
	requestBody = {'apiVersion': 'v1', 'dryRun': ['All'], 'gracePeriodSeconds': 0, 'kind': 'Pod', 'orphanDependents': True, 'propagationPolicy': 'Orphan'}
	"""

	header = {
		'accept': 'application/json',
		'content-type': 'application/x-www-form-urlencoded'
	}
	queryParam = {'dryRun' : dryRun, 'gracePeriodSeconds' : gracePeriodSeconds, 'orphanDependents' : orphanDependents, 'propagationPolicy' : propagationPolicy}

	api_url = f"http://xxxx:8080/api/v1/namespaces/{namespace}/limitranges/{name}"
	response = requests.delete(api_url, headers=header, params=queryParam, json=requestBody)
	return {"status_code": response.status_code, "response": response.json()}

\end{minted}
\caption{Enrich-3 Python tool}\label{fig:enrich3-tool}
\end{figure*}

\end{document}